\documentclass[12pt]{article}

\pdfoutput=1
 
\usepackage{amsmath,amssymb,amsfonts,amssymb}
\usepackage{eucal}
\usepackage{mathrsfs}
\usepackage{braket}
\usepackage{slashed}
\usepackage{graphicx}
\usepackage{bm}
\usepackage{cite}
\usepackage{tikz}
\usepackage[colorinlistoftodos]{todonotes}
\usepackage{subcaption}
\usepackage[hidelinks]{hyperref}

\hypersetup{
    colorlinks=true,
    linkcolor=black,
    citecolor=black,
    filecolor=black,
    urlcolor=black,
}

\usetikzlibrary{decorations.pathmorphing, calc}
\tikzset{snake it/.style={decorate, decoration=snake}}
\usepackage{titling}

\newcommand{\mb}{\mathbf}

\makeatletter
\newsavebox{\@brx}
\newcommand{\llangle}[1][]{\savebox{\@brx}{\(\m@th{#1\langle}\)}%
  \mathopen{\copy\@brx\kern-0.5\wd\@brx\usebox{\@brx}}}
\newcommand{\rrangle}[1][]{\savebox{\@brx}{\(\m@th{#1\rangle}\)}%
  \mathclose{\copy\@brx\kern-0.51\wd\@brx\usebox{\@brx}}}
\makeatother

\usepackage{titling}
\usepackage{authblk}

\title{On the need for soft dressing}

\author[1,2]{Daniel Carney\thanks{carney@umd.edu}}
\author[3]{Laurent Chaurette\thanks{dodeca@phas.ubc.ca}}
\author[3]{Dominik Neuenfeld\thanks{dneuenfe@phas.ubc.ca}}
\author[3]{Gordon Semenoff\thanks{gordonws@phas.ubc.ca}}
\affil[1]{Joint Center for Quantum Information and Computer Science, University of Maryland, \mbox{College Park, MD 20742 USA} }
\affil[2]{Joint Quantum Institute, National Institute of Standards and Technology, \mbox{Gaithersburg, MD 20899 USA} }
\affil[3]{Department of Physics and Astronomy, University of British Columbia, \mbox{Vancouver, BC V6T 1Z1 Canada}}

\begin{document}

\maketitle

\begin{abstract}
In order to deal with IR divergences arising in QED or perturbative quantum gravity scattering processes, one can either calculate inclusive quantities or use dressed asymptotic states. We consider incoming superpositions of momentum eigenstates and show that in calculations of cross-sections these two approaches yield different answers: in the inclusive formalism no interference occurs for incoming finite superpositions and wavepackets do not scatter at all, while the dressed formalism yields the expected interference terms. This suggests that rather than Fock space states, one should use Faddeev-Kulish-type dressed states to correctly describe physical processes involving incoming superpositions. We interpret this in terms of selection rules due to large $U(1)$ gauge symmetries and BMS supertranslations.
\end{abstract}

\newpage
\setcounter{tocdepth}{1}

\section{Introduction}

Quantum electrodynamics and perturbative quantum gravity are effective quantum field theories
which describe the two long-ranged forces seen in nature. They also both suffer from infrared divergences coming from virtual boson loops in Feynman diagrams in the perturbative computation of the S-matrix. These divergences exponentiate when resummed and set the amplitude for any process between a finite number of interacting particles to zero. This is known as the infrared catastrophe.

One proposed resolution of the infrared catastrophe is to consider only inclusive quantities, for example soft-inclusive transition probabilities in the context of scattering theory, which are defined by summing over the production of any number of soft photons and gravitons.
In the case of electrodynamics, this resolution dates back to Bloch and Nordsieck \cite{Bloch1937,Yennie1961} and, in perturbative quantum gravity, it was developed by Weinberg \cite{Weinberg1965}. The contributions from emitted soft bosons cancel the IR divergences from the virtual loops. An upshot of this solution of the infrared problem is the fact that, in QED, any non-trivial scattering process involving charged particles inevitably produces a cloud of an infinite number of arbitrarily soft photons. In the case of quantum gravity, soft gravitons are produced, and, since all particles carry gravitational charge, IR divergences arise in any scattering process. The use of inclusive probabilities is justified by the assumption that the softest photons and gravitons must escape detection. These bosons carry very little energy and have a negligible effect on the kinematics of the process. However, it was recently shown that they carry a lot of information in the sense that their quantum states are highly entangled with those of the charged particles. The loss of the soft particles results in decoherence of the final state of the hard particles, where the momentum eigenstates for electrically or gravitationally charged particles are the pointer basis \cite{Carney2017,Zurek:1981xq}. See refs. \cite{Breuer2001,Calucci2003,Calucci2004} for related work.

The infrared catastrophe can be traced back to the long-ranged nature of the interactions which is in conflict with the assumption of asymptotic decoupling needed to formulate scattering theory \cite{dollard}. An approach to the infrared problem, alternative to using inclusive probabilities, is to use \emph{dressed} states which are defined by including the aforementioned clouds of soft photons and gravitons with the asymptotic states \cite{Chung1965,Kibble1968a,Kibble1968b,Kibble1968c,Kibble1968d,Kulish1970,Zwanziger:1974jz,Ware2013ConstructionGravity,Dybalski2017}. Faddeev and Kulish argued that such an approach diagonalizes the correct asymptotic Hamiltonian and therefore yields the asymptotic decoupling which is necessary for a satisfactory formulation of scattering theory. The detailed structure of the coherent states can be adjusted so as to cancel the infrared divergences in the S-matrix, providing an IR-finite S-matrix and scattering probabilities. However, the out-going states still contain particles accompanied by soft photon and graviton clouds. One can ask the same question: given these infrared safe states, what is the nature of the state of the outgoing hard particles? The answer is that precisely the same decoherence is found to occur in either the inclusive or dressed approaches \cite{Carney2017a}, i.e.~there is still a lot of information in the entanglement between the hard particles and the radiation.\footnote{Note, there are also other proposals for how to define an IR finite density matrix \cite{Gomez2017InfraredCoherence}, which we will not discuss here.}

Both the dressed and inclusive formalisms are designed to give the same predictions for the probability of scattering from an incoming set of momenta $\mb{p}_1, \ldots, \mb{p}_n$ into an outgoing set of momenta $\mb{p}'_1, \ldots, \mb{p}'_m$. The measurement of observables which only depend on the hard particles should be predictable from the reduced density matrix obtained by tracing over soft bosons, which are invisible to a finite size detector. Given an incoming momentum eigenstate the two formalisms agree. Thus, one might naively think for calculating cross-sections it does not matter which formalism one chooses. We show in this paper that this is not the case: the two approaches differ in their treatment of incoming superpositions. Consider a simple superposition of two momentum eigenstates for a single charged particle
\begin{align}
\ket \psi = \frac{1}{\sqrt 2} (\ket{\mb{p}} + \ket{\mb{q}}),
\end{align}
scattering off of a classical potential. We expect the out-state to be described by a density matrix of the form
\begin{align}
\label{eq:out_density_matrix_expected}
\rho = \frac 1 2 S \left(\ket{\mb{p}}\bra{\mb{p}} + \ket{\mb{p}}\bra{\mb{q}} + \ket{\mb{q}}\bra{\mb{p}} + \ket{\mb{q}}\bra{\mb{q}} \right) S^\dagger.
\end{align}
Here $S$ is the scattering operator and we have performed a trace over the soft radiation, hence $\rho$ is the density matrix for the hard particles. If $\ket{\mb{p}}, \ket{\mb{q}}$ are correctly dressed states, this expectation is indeed correct. In the inclusive formalism, however, where $\ket{\mb{p}}$, $\ket{\mb{q}}$ are Fock space momentum eigenstates, there is no interference between the different momenta as opposed to the diagonal terms of \eqref{eq:out_density_matrix_expected}. We find that the diagonal entries of the density matrix which encode the cross-sections are of the form
\begin{align}
\sigma_{\psi \to out} \propto \bra{out} \rho^{\text{incl}} \ket{out} = \frac 1 2 \bra{out}\ S \left(\ket{\mb{p}}\bra{\mb{p}} + \ket{\mb{q}}\bra{\mb{q}} \right) S^\dagger \ket{out}.
\end{align}
In other words, the cross-section behaves as if we had started with a classical ensemble of states with momenta $\mb{p}$ and $\mb{q}$. The entire scattering history is decohered by the loss of the soft radiation. This appears to contrast starkly with any realistic experiment.

Moreover, as we will show, repeating the analysis for wavepackets, e.g. $\ket \psi = \int d\mb{p} f(\mb{p}) \ket {\mb{p}}$, leads to the nonsensical conclusion that a wave-packet is not observed to scatter at all. However, in the dressed state formalism of Faddeev-Kulish the interference appears as in equation \eqref{eq:out_density_matrix_expected}. This strongly suggests that scattering theory in quantum electrodynamics and perturbative quantum gravity should really not be formulated in terms of standard Fock states of charged particles. Formulating the theories using dressed states seems to be a good alternative.

Dressed states also arise naturally in the recent discussions of asymptotic gauge symmetries \cite{Strominger2014OnScattering,Kapec2015,Campiglia2015,CampigliaMiguelandLaddha2015AsymptoticParticles,Hawking:2016sgy,Strominger2018LecturesTheory}, which imply the existence of selection sectors\cite{Gabai2016LargeQED,Kapec2017,Choi2017,Choi2018AsymptoticSupertranslations}. See also \cite{Afshar:2016uax,Mishra:2017zan} for work on soft charges and dressing in holography. Our findings have a nice interpretation in the language of this program: only superpositions of states within the same selection sector can interfere. This explains the failure of the undressed approach. In the inclusive formalism, essentially any pair of momentum eigenstates live in different charge sectors. In contrast, the Faddeev-Kulish formalism is designed so that all of the dressed states live within the same charge sector. 

Our results can also be viewed in the context of the black hole information problem \cite{Hawking1975ParticleHoles,Almheiri2013BlackFirewalls}. In particular, Hawking, Perry, and Strominger \cite{Hawking2016} and Strominger \cite{Strominger2017} have recently suggested that black hole information may be encoded in soft radiation. In black hole thought experiments, one typically imagines preparing an initial state of wavepackets organized to scatter with high probability to form an intermediate black hole. Our results suggest then that one needs to use dressed initial states to study this problem. See also \cite{Mirbabayi2016,Bousso:2017dny} for some remarks on the use of dressed or inclusive formalisms for studying black hole information.

The rest of the paper is organized as follows. We start by presenting the calculations showing that the dressed and undressed formalism disagree in section \ref{sec:discrete} for discrete superpositions and in section \ref{sec:continuous} for wavepackets. The discussion and interpretation of the results takes place in section \ref{sec:implications}. There, we will argue why our findings imply that dressed states are better suited to describe scattering than the inclusive Fock-space formalism. We will give a new very short argument for the known result of \cite{Choi2017} that the dressing operators and the S-matrix weakly commute and argue for a more general form of dressing beyond Faddeev-Kulish. We will then interpret our results in terms of asymptotic symmetries and selection sectors before concluding in section \ref{sec:conclusions}. The appendix contains proofs of certain statements in sections \ref{sec:discrete} and \ref{sec:continuous}.

\section{Scattering of discrete superpositions}
\label{sec:discrete}
In this and the next section we generalize the results of \cite{Carney2017} to the case of incoming superpositions of momentum eigenstates. We begin in this section by studying discrete superpositions $\ket{\psi} = \ket{\alpha_1} + \cdots + \ket{\alpha_N}$ of states with various momenta $\alpha = \mb{p}_1, \mb{p}_2, \ldots$. We will see that the dressed and inclusive formalisms give vastly different predictions for the probability distribution of the outgoing momenta: dressed states will exhibit interference between the $\alpha_i$ whereas undressed states do not. 

\subsection{Inclusive formalism}
Consider scattering of an incoming superposition of charged momentum eigenstates 
\begin{align}
\ket{in} = \sum_i^N f_i \ket{\alpha_i},
\end{align}
with $\sum_i |f_i|^2 = 1$. The outgoing density matrix vanishes due to IR divergences in virtual photon loops. However, we can obtain a finite result if we trace over outgoing radiation \cite{Bloch1937,Weinberg1965,Yennie1961,Carney2017}. The resulting reduced density matrix of the hard particles takes the form
\begin{align}
\label{eq:scattering_density_matrix}
\rho = \sum_{b}\sum_{i,j}^N \;\iint d\beta\;d\beta'f_if^{*}_j S_{\beta b,\alpha_i}S^{*}_{\beta'b,\alpha_j}\ket{\beta}\bra{\beta'},
\end{align}
where $\beta$ and $\beta'$ are lists of the momenta of hard particles in the outgoing state, and the sum over $b$ denotes the trace over soft bosons. We will be interested in the effect of infrared divergences on this expression.

The sum over external soft boson states $b$ produces IR divergences which cancel those coming from virtual boson loops. We can regulate these divergences by introducing an IR cutoff (e.g.~a soft boson mass $\lambda$). Following the standard soft photon resummation techniques \cite{Weinberg1965}, one finds that the total effect of these divergences yields reduced density matrix elements of the form
\begin{align}
\label{dmfirstresult}
\rho_{\beta\beta'} = \sum_{i,j}^N f_i f^{*}_j S^{\Lambda}_{\beta \alpha_i}S^{\Lambda*}_{\beta'\alpha_j} \lambda^{\Delta A_{\beta\beta',\alpha_i \alpha_j}+\Delta B_{\beta\beta',\alpha_i \alpha_j}} \mathcal{F}_{\beta\beta',\alpha_i\alpha_j}(E,E_T,\Lambda).
\end{align}
Here we have introduced ``UV'' cutoffs $\Lambda,E$ on the virtual and real soft boson energies, so $S^{\Lambda}$ are $S$-matrix elements with the soft boson loops cut off below $\Lambda$ and we only trace over outgoing bosons with individual energies up to $E$ and total energy $E_T$. The explicit form of the Sudakov rescaling function $\mathcal{F}$ can be found in \cite{Carney2017}. What concerns us here is the behavior of this expression in the limit where we remove the IR regulator $\lambda \to 0$, which is controlled by the exponents
\begin{align}
\label{eq:def_exponents}
\begin{split}
\Delta A_{\beta\beta',\alpha\alpha'} & =
-\frac{1}{2}\sum_{n,n'\in\alpha,\bar{\alpha}',\beta,\bar{\beta}'}\frac{e_{n}e_{n'}\eta_{n}\eta_{n'}}{8\pi^{2}}\beta^{-1}_{nn'}\ln\left[\frac{1+\beta_{nn'}}{1-\beta_{nn'}}\right], \\
\Delta B_{\beta\beta',\alpha\alpha'} & =
-\frac{1}{2}\sum_{n,n'\in\alpha,\bar{\alpha}',\beta,\bar{\beta}'}\frac{m_{n}m_{n'}\eta_{n}\eta_{n'}}{16\pi^{2}M_{p}^{2}}\beta^{-1}_{nn'}\frac{1+\beta_{nn'}^{2}}{\sqrt{1-\beta_{nn'}^{2}}}\ln\left[\frac{1+\beta_{nn'}}{1-\beta_{nn'}}\right].
\end{split}
\end{align}
The factor $\eta_{n}$ is defined as $+1$ $(-1)$ if particle n is incoming (outgoing). The quantities $\beta_{nn'} = \sqrt{1-\frac{m_{n}^{2}m_{n'}^{2}}{(p_{n}\cdot p_{n'})^{2}}}$ are the relative velocities between pairs of particles and a bar interchanges incoming states for outgoing and vice versa. The expressions for $\Delta A$ and $\Delta B$ come from contributions of soft photons and gravitons, respectively. The question now is which terms survive.

The special case of no superposition, $\alpha_i = \alpha_j = \alpha$, was discussed in \cite{Carney2017}. There it was shown that $\Delta A_{\beta\beta',\alpha\alpha} \geq 0$ and $\Delta B_{\beta\beta',\alpha\alpha} \geq 0$, so that in the limit $\lambda \to 0$, all of the terms in the sum except those with $\Delta A = \Delta B = 0$ will vanish. The equality holds if and only if the out states $\beta$ and $\beta'$ contain particles such that the amount of electrical charge and mass carried with any choice of velocity agrees for $\beta$ and $\beta'$. This can be phrased in terms of an infinite set of operators which measure charges flowing along a velocity $\mathbf v$. These are defined as
\begin{align}
\begin{split}
\label{currentdefs}
 \hat j^{em}_{\mathbf v} &= \sum_i e_i a_{i,\mathbf p_i(\mathbf v) }^{\dagger} a_{i,\mathbf p_i(\mathbf v) },\\
 \hat j^{gr}_{\mathbf v} &= \sum_i E_i(\mathbf v) a_{i,\mathbf p_i(\mathbf v) }^{\dagger} a_{i,\mathbf p_i(\mathbf v) }, \\
 \hat j^{gr, 0}_{\mathbf v} &= \sum_i\int d \omega \, \, \omega a_{i, \mathbf v \omega}^{ \dagger} a_{i,\mathbf v \omega},
 \end{split}
 \end{align}
for charged particles, massive particles and hard massless particles, respectively. The sum runs over all particle species. Clearly, momentum eigenstates are also eigenstates of these operators. Using these operators, the equality of currents can be expressed as
\begin{align}
\label{eq:decoherence_condition_no_superposition}
\hat{j}_{\textbf{v}}\ket{\beta} \sim \hat{j}_{\textbf{v}}\ket{\beta'},
\end{align}
where the tilde means that the eigenvalues of the states are the same on both sides for all velocities. In appendix \ref{app:proof_positivity}, we show that the more general exponents $\Delta A_{\beta\beta',\alpha \alpha'}$ and $\Delta B_{\beta\beta',\alpha \alpha'}$ are positive. Similarly to the argument in \cite{Carney2017a}, one can show that $\Delta A$ and $\Delta B$ are non-zero if and only if
\begin{align}
\label{eq:QED_wavepacket_decoherence}
\hat{j}_{\textbf{v}}\ket{\alpha_i} +\hat{j}_{\textbf{v}}\ket{\beta'} \sim \hat{j}_{\textbf{v}}\ket{\alpha_j}+\hat{j}_{\textbf{v}}\ket{\beta},
\end{align}
that is if the list of hard currents in states $\ket{\alpha}$ and $\ket{\beta'}$ is the same as the list of hard currents in states $\ket{\alpha'}$ and $\ket{\beta}$. An easy way to understand the form of equation \eqref{eq:QED_wavepacket_decoherence} is by looking at equation \eqref{eq:def_exponents}. There, the bar over $\alpha'$ (which corresponds to $\alpha_j$) indicates that it should be treated as an outgoing particle, i.e. similarly to $\beta$. On the other hand $\bar \beta'$ should be treated similarly to $\alpha$. Hence, we obtain equation \eqref{eq:QED_wavepacket_decoherence} from \eqref{eq:decoherence_condition_no_superposition} by replacing $\alpha_i \to \alpha_i + \beta'$ and $\alpha_j \to \alpha_j + \beta$. On the other hand it is clear that in the case of $\ket {\alpha_i} = \ket{\alpha_j} = \ket \alpha$ equation \eqref{eq:QED_wavepacket_decoherence} reduces to equation \eqref{eq:decoherence_condition_no_superposition}.

Armed with these results, we can calculate the cross-sections given an incoming superposition. These are proportional to the diagonal elements $\beta = \beta'$ of the density matrix; for simplicity we ignore forward scattering terms. The diagonal terms of the density matrix \eqref{dmfirstresult} are proportional to $\lambda^{\Delta A+\Delta B}$. This factor reduces to unity if $\hat j_{\mathbf v} \ket{\alpha_i} \sim \hat j_{\mathbf v} \ket{\alpha_j}$ for all of the currents \eqref{currentdefs} and is zero otherwise. For a generic superposition, this implies that only terms with $i=j$ contribute and we find 
\begin{align}
\label{discreteanswerundressed}
\sigma_{\text{in} \to \beta} \propto \rho_{\beta\beta} = \sum_{i,j}^N f_i f_j^* \mathcal{F}_{\beta\beta,\alpha_i\alpha_j} S_{\beta\alpha_i}^\Lambda S_{\beta\alpha_j}^{\Lambda*} \delta_{\alpha_i \alpha_j}= \sum_i^N |f_i|^2 |S_{\beta,\alpha_i}^\Lambda|^2 \mathcal{F}_{\beta\beta,\alpha_i\alpha_i} .
\end{align}
As we see, no interference terms between incoming states are present. Instead, the total cross-section is calculated as if the incoming states were part of a classical ensemble with probabilities $|f_i\mathcal|^2$. The reason is that in the inclusive approach the information about the interference is carried away by unobservable soft radiation. To define the scattering cross-section, however, we need to trace out the soft radiation and we obtain the above prediction, which is at odds with the naive expectation, equation \eqref{eq:out_density_matrix_expected}.

\subsection{Dressed formalism}
The calculation above was done using the usual, undressed Fock states of hard charges, which required to calculate inclusive cross-sections. The alternative approach we will now turn to is to consider transitions between dressed states. For concreteness, we will follow the dressing approach of Chung and Faddeev-Kulish\footnote{Recently, a generalization of Faddeev-Kulish states was suggested \cite{Kapec2017}. We will extend our discussion to those states in section \ref{sec:implications}.}, which contains charged particles accompanied by a cloud of real bosons which radiate out to lightlike infinity \cite{Kulish1970,Chung1965,Ware2013ConstructionGravity}. For a given set of momenta $\alpha = \mb{p}_1, \mb{p}_2, \ldots$, we write the dressed state as\footnote{The double bracket notation is due to \cite{Mirbabayi2016}. The previous paper of the authors \cite{Carney2017a} used $\ket{\tilde \alpha}$ to denote dressed states. The authors regret this life decision.}
\begin{align}
\label{eq:def_dressing}
 \| \alpha \rrangle \equiv W_\alpha \ket \alpha.
\end{align}
The operator $W_\alpha$ equips the state $\ket \alpha$ with a cloud of photons/gravitons. 
For QED, $W_\alpha$ is the unitary operator (with a finite IR cutoff $\lambda$)
\begin{align}\label{eq:Dressing}
 W_\alpha \equiv \exp \left\{ e \sum_{l=1}^2 \int_\lambda^{E} \frac{d^3 \mathbf k}{\sqrt{2 k }} \left( F_l(\mathbf k , \alpha)  a_{\mathbf k}^{l\dagger} - F^*_l(\mathbf k , \alpha)  a_{\mathbf k}^{l} \right) \right\},
\end{align}
where $a_{\mathbf k}^{l \dagger}$ creates a photon in the polarization state $l$ and the soft factor
 \begin{align}
F_l(\mathbf k , \alpha) = \sum_{\mathbf p \in \alpha}\frac{\epsilon_l \cdot  p}{k \cdot p} \phi(\mathbf k, \mathbf p)
 \end{align}
depends on the polarization vectors $\epsilon_l$ and some smooth, real function $\phi(\mathbf k, \mathbf p)$ which goes to $1$ as $|\mathbf k| \to 0$. Letting $W$ act on Fock space states for $\lambda = 0$ gives states with vanishing normalization, hence in the strict $\lambda \to 0$ limit $W$ is no good operator on Fock space. Thus, as before, we will do calculations with finite $\lambda$ and only at the end we will take $\lambda \to 0$.\footnote{Note that as argued in \cite{Kulish1970}, a proper definition of $W$ in the limit $\lambda \to 0$ should be possible on a von Neumann space. }

The Faddeev-Kulish construction was adapted to perturbative quantum gravity in \cite{Ware2013ConstructionGravity}. In this case the dressing has the same form as equation \eqref{eq:Dressing}, the only difference being that $a$ ($a^\dagger$) is now a graviton annihilation (creation) operator and the functions $F$ depend on the polarization tensor $\epsilon_{\mu\nu}$ \cite{Ware2013ConstructionGravity},
\begin{align}
F^{gr}_l(\mathbf k , \alpha) = \sum_{\mathbf p \in \alpha} \frac{p_\mu \epsilon_l^{\mu\nu} p_\nu}{k \cdot p} \phi(\mathbf k, \mathbf p).
\end{align}
S-matrix elements taken between dressed states
\begin{align}
   \mathbb{S}_{\beta \alpha} \equiv \llangle \beta \| S \| \alpha \rrangle = \bra{\beta} W_\beta^\dagger S W_\alpha \ket \alpha
\end{align}
are independent of $\lambda$ and thus finite as $\lambda \to 0$. The Sudakov factor $\mathcal F$ is contained in the dressed S-matrix elements.\footnote{The actual definition of the S-matrix should also contain a term to cancel the infinite Coulomb phase factor. Since this is immaterial to the current discussion we neglect this subtlety.}

Consider now an incoming state consisting of a discrete superposition of such dressed states,
\begin{align}
\| in \rrangle = \sum_i f_i \| \alpha_i \rrangle.
\end{align}
The outgoing density matrix is then
\begin{align}
\rho = \sum_{i,j} \iint d\beta d\beta' f_i f^*_j \mathbb{S}_{\beta \alpha_i} \mathbb{S}_{\beta' \alpha_j}^* \|\beta \rrangle \llangle \beta' \|.
\end{align}
This density matrix is formally unitary, however, every experiment should be able to ignore soft radiation. Following \cite{Carney2017}, we treat the soft modes as unobservable and trace them out. This yields the reduced density matrix for the outgoing hard particles,
\begin{align}
\label{dressedoutdm}
\rho^{hard}_{\beta\beta'} = \sum_{i,j} f_i f^*_j \mathbb{S}_{\beta \alpha_i} \mathbb{S}_{\beta' \alpha_j}^* \braket{ 0 | W_{\beta}^{\dagger} W_{\beta'} | 0}.
\end{align}
The last term is the photon vacuum expectation value of the out-state dressing operators. This factor reduces to one or zero as shown in \cite{Carney2017}; one if $\hat j(\beta) \sim \hat j(\beta')$ and zero otherwise. This is responsible for the decay of most off-diagonal elements in \eqref{dressedoutdm}. However, if we are interested in the cross-section for a particular outgoing state $\beta$, this is again given by a diagonal density matrix element,
\begin{align}
\label{discreteanswerdressed}
\sigma_{\text{in} \to \beta} \propto \rho_{\beta\beta} = \sum_{i,j} f_i f^*_j \mathbb{S}_{\beta \alpha_i} \mathbb{S}_{\beta \alpha_j}^*.
\end{align}
In stark contrast to the result obtained in the inclusive formalism, equation \eqref{discreteanswerundressed}, this cross-section exhibits the usual interference between the various incoming states, c.f.~equation \eqref{eq:out_density_matrix_expected}. The reason for this is that in the dressed formalism, the outgoing radiation is described by the dressing which only depends on the out-state and not on the in-state. We will discuss this in more detail in section \ref{sec:implications}. This establishes that the inclusive and dressed formalism are not equivalent but yield different predictions for cross-sections of finite superpositions.

\section{Wavepackets}
\label{sec:continuous}
We will now proceed to look at scattering of wavepackets and find that the result is even more disturbing. After tracing out infrared radiation in the undressed formalism, no indication of scattering is left in the hard system. On the contrary, once again we will see that with dressed states, one gets the expected scattering out-state.

\subsection{Inclusive formalism}
We consider incoming wavepackets of the form
\begin{align}
\ket{in} = \int d\alpha f(\alpha) \ket{\alpha},
\end{align}
normalized such that $\int d\alpha |f(\alpha)|^2 = 1$. The full analysis of the preceding section still applies, provided we replace $\sum_{\alpha_i} \to \int d\alpha$, $\alpha_i \to \alpha$, $f_i \to f(\alpha)$ and similarly for $a_j \to \alpha'$. The only notable exception is the calculation of single matrix elements as in equation \eqref{discreteanswerundressed}, which now reads
\begin{align}
\rho_{\beta\beta} = \iint d \alpha d \alpha'  f(\alpha) f^*(\alpha') S_{\beta,\alpha}^\Lambda S_{\beta,\alpha'}^{\Lambda*} \delta_{\alpha \alpha'} \mathcal{F}_{\beta\beta,\alpha\alpha'}(E,E_T,\Lambda).
\end{align}
Note that here, by the same argument as before, the $\lambda$-dependent factor is turned into a Kronecker delta, which now reduces the integrand to a measure zero subset on the domain of integration. The only term that survives the integration is the initial state, which is acted on with the usual Dirac delta $\delta(\alpha-\beta)$, i.e. the ``$1$'' term in $S = 1 - 2 \pi i \mathcal M$. The detailed argument can be found in appendix \ref{app:checks}. Thus we conclude that
\begin{align}
\rho^{out}_{\beta\beta'} = f(\beta) f^*(\beta') = \rho^{in}_{\beta\beta'}.
\end{align}
The hard particles show no sign of a scattering event.

\subsection{Dressed wavepackets}
The dressed formalism has perfectly reasonable scattering behavior. Consider wavepackets built from dressed states
\begin{equation}
\| in \rrangle = \int d\alpha\;f(\alpha) \| \alpha \rrangle,
\end{equation}
with $\| \alpha \rrangle$ a dressed state in the same notation as in equation \eqref{eq:def_dressing}. The S-matrix applied on dressed states is infrared-finite and the outgoing density matrix can be expressed as
\begin{align}
\rho = \iint d\beta d\beta' \iint d\alpha d\alpha' f(\alpha)f^{*}(\alpha')  \mathbb{S}_{\beta \alpha}\mathbb{S}^*_{\beta'\alpha'} \|\mathbf \beta \rrangle \llangle \mathbf \beta' \|.
\end{align}
Tracing over soft modes, we find
\begin{align}
\rho_{\beta\beta'} = \iint d\alpha d\alpha'  f(\alpha) f^*(\alpha')\mathbb{S}_{\beta \alpha}\mathbb{S}^*_{\beta'\alpha'} \braket{ W^{\dagger}_{\beta} W_{\beta'} }.
\end{align}
Again the expectation value is taken in the photon vacuum. The crucial point here is that this factor is independent of the initial states $\alpha$. Upon sending the IR cutoff $\lambda$ to zero, the expectation value for $W^{\dagger} W$ takes only the values 1 or 0, leading to decoherence in the outgoing state, but the cross-sections still exhibit all the usual interference between components of the incoming wavefunction,
\begin{align}
\rho_{\beta\beta} = \iint d\alpha d\alpha'  f(\alpha) f^*(\alpha')\mathbb{S}_{\beta \alpha}\mathbb{S}^*_{\beta\alpha'},
\end{align}
unlike in the inclusive formalism.

\section{Implications}
\label{sec:implications}
In this section we will discuss the implications of our results and generalize and re-interpret our findings in particular in view of asymptotic gauge symmetries in QED and perturbative quantum gravity.

\subsection{Physical interpretation}
The reason for the different predictions of the inclusive and dressed formalism is the IR radiation produced in the scattering process. The key idea is that accelerated charges produce radiation fields made from soft bosons. In the far infrared, the radiation spectrum has poles as the photon frequency $k^0 \to 0$ of the form $p_i/p_i \cdot k$, where $p_i$ are the hard momenta. These poles reflect the fact that the radiation states are essentially classical and are completely distinguishable for different sets of asymptotic currents $\hat j_{\mb{v}}$.

In the inclusive formalism, we imagine incoming states with no radiation, and so the outgoing radiation state has poles from both the incoming hard particles $\alpha$ and the outgoing hard particles $\beta$. In the dressed formalism, the incoming part of the radiation is instead folded into the dressed state $\| \alpha \rrangle$, which in the Faddeev-Kulish approach is designed precisely so that the outgoing radiation field \emph{only} includes the poles from the outgoing hard particles. Thus if we scatter undressed Fock space states, a measurement of the radiation field at late times would completely determine the entire dynamical history of the process $\alpha \to \beta$, leading to the classical answer \eqref{discreteanswerundressed}. If we instead scatter dressed states, the outgoing radiation has incomplete information about the incoming charged state, which is why the various incoming states still interfere in \eqref{discreteanswerdressed}. Given that this type of interference is observed all the time in nature, this seems to strongly suggest that the dressed formalism is correct for any problem involving incoming superpositions of momenta.

Based on the result of section \ref{sec:discrete}, one might argue that equation \eqref{discreteanswerundressed} perhaps is the correct answer and one would have to test experimentally whether or not interference terms appear if we give a scattering process enough time so that the decoherence becomes sizable. After all, the inclusive and dressed approach to calculating cross-sections are at least in principle distinguishable, although maybe not in practice due to very long decoherence times. However, we have demonstrated in section \ref{sec:continuous} that the inclusive formalism predicts an even more problematic result for continuous superpositions, namely that no scattering is observed at all. We thus propose that using the dressed formalism is the most conservative and physically sensible solution to the problem of vanishing interference presented in this paper.

\subsection{Allowed dressings}
\subsubsection*{Dressing operators weakly commute with the S-matrix}
It was conjectured in \cite{Kapec2017} and proven in \cite{Choi2017} that the far IR part of the dressing weakly commutes with the S-matrix to leading order in the energy of the bosons contained in the dressing. In particular, this means that the amplitudes
\begin{align}
   \bra \beta W^\dagger_\beta S W_\alpha \ket \alpha \sim \bra \beta W^\dagger_\beta W_\alpha S \ket \alpha  \sim \bra \beta S W^\dagger_\beta W_\alpha  \ket \alpha
\end{align}
are all IR finite, while they might differ by a finite amount. A short proof of this in QED, complementary to \cite{Choi2017}, can be given as follows (the gravitational case follows analogously). Recall that Weinberg's soft theorem for QED states that to lowest order in the soft photon momentum $\mathbf q$ of outgoing soft photons
\begin{align}
\braket{\epsilon_{l_1} a_{\mathbf q_1}^{l_1}\dots \epsilon_{l_N}  a_{\mathbf q_N}^{l_N} S } \sim \prod_{i=1}^N \left( \sum_j^M \eta_j e_j  \frac{\epsilon_{l_i} \cdot p_j}{q_i \cdot p_j} \right)\braket{S}.
\end{align}
A similar argument holds for incoming photons. For incoming photons with momentum $\mathbf q$ we find that
\begin{align}
\braket{S \epsilon^*_{l_1} a^{l_1 \dagger}_{\mathbf q_1}\dots \epsilon^*_{l_N} a^{l_N \dagger }_{\mathbf q_N}} \sim \prod_{i=1}^N \left( - \sum_j^M \eta_j e_j  \frac{\epsilon^*_{l_i} \cdot p_j}{q_i \cdot p_j} \right)\braket{S}.
\end{align}
The reason for the relative minus sign is that incoming photons add energy-momentum to lines in the diagram instead of removing it. That means that the momentum in the denominator of the propagator changes $(p-q)^2 + m^2 \to (p+q) + m^2$ and vice versa. For small momentum, the denominator becomes $-2pq \to 2pq$. From this it directly follows that for general dressings at leading order in the IR divergences,
\begin{align}
\begin{split}
\braket{SW} = \braket{ S e^{ \int d^3 k  ( F_l(\mathbf k) a_{\mathbf k}^{l\dagger} - F^*_l(\mathbf k) a_{\mathbf k}^l)}} & \sim  \mathcal N \braket{ S e^{ \int d^3 k F_l(\mathbf k) a_{\mathbf k}^{l\dagger}}} \\
& \sim \mathcal N \braket{ e^{- \int d^3 k F^*_l(\mathbf k) a_{\mathbf k}^{l}} S }\\ & \sim \braket{ e^{ \int d^3 k  ( F_l(\mathbf k) a_{\mathbf k}^{l\dagger} - F^*_l(\mathbf k) a_{\mathbf k}^l)} S} = \braket{WS}.
\end{split}
\end{align}
In the first and third step we have split the exponential using the Baker-Campbell-Hausdorff formula ($\mathcal N$ is the normalization which is finite for finite $\lambda$) and in the second equality we have used Weinberg's soft theorem for outgoing and incoming particles. 

\subsubsection*{Dressings cannot be arbitrarily moved between in- and out-states}
This opens up the question about the most general structure of a consistent Faddeev-Kulish-like dressing. For example, one could ask whether one can consistently define S-matrix elements with the dressing only acting on the out-state. To answer this question, we assume that the dressing of the out-state has the same IR structure as equation \eqref{eq:Dressing}, but is more general in that it may also include the momenta of (some) particles of the in-state, i.e. $W_{\beta} \to W_{\beta} W_{\tilde \alpha}$ or any other momenta which might not even appear in the process, $W_{\beta} W_{\tilde \alpha} \to W_{\beta} W_{\tilde \alpha} W_{\zeta} $. The IR structure of the in-dressing is then fixed by the requirement that the S-matrix element is finite. In addition to the requirement of IR-finiteness we ask that the so defined S-matrix elements give rise to the correct rules for superposition and the correct scattering for wavepackets, even after tracing out soft radiation.

Applying the logic of the previous sections and \cite{Carney2017a}, one finds that tracing over the soft bosons yields for a diagonal matrix element $\rho_{\beta\beta}$
\begin{align}
\rho^{hard}_{\beta\beta} = \sum_{i,j} f_i f^*_j \mathbb{S}_{\beta \alpha_i} \mathbb{S}_{\beta' \alpha'_j}^* \braket{ 0 | W_{\tilde \alpha'}^{\dagger} W_{\tilde \alpha} | 0}
\end{align}
and
\begin{align}
\rho^{hard}_{\beta\beta} =  \iint d\alpha d\alpha' f(\alpha)f^{*}(\alpha')  \mathbb{S}_{\beta \alpha}\mathbb{S}^*_{\beta'\alpha'} \braket{ 0 | W_{\tilde \alpha'}^{\dagger} W_{\tilde \alpha} | 0}
\end{align}
for finite and continuous superpositions, respectively.
Here, we have used that
\begin{align}
\left.\braket {W_{\tilde \alpha'}^\dagger W_{\beta'}^\dagger W_{\beta}W_{\tilde \alpha} } \right|_{\beta = \beta'} = \braket {W_{\tilde \alpha'}^\dagger W_{\tilde \alpha}}.
\end{align}
The expectation value is taken in the soft boson Fock space. The expression in the case of $\tilde \alpha = \alpha$ and $\tilde \alpha' = \alpha'$ was already encountered in sections \ref{sec:discrete} and \ref{sec:continuous} in the context of inclusive calculations, where it was responsible for the unphysical form of the cross-sections. By the same logic it follows that even in the case where $\tilde \alpha$ is a proper subset of $\alpha$, we will obtain a Kronecker delta which sets $\tilde \alpha = \tilde \alpha'$ and we again do not obtain the expected form of the cross-section. Instead, particles from the subset $\tilde \alpha$ will cease to interfere. We thus conclude that the dressing of the out-states must be independent of the in-states and it is not consistent to build superposition of states which are dressed differently. This means that building superpositions from hard and charged Fock space states is not meaningful. In particular, we cannot use undressed states to span the in-state space by simply moving all dressings to the out-state.

\subsubsection*{Generalized Faddeev-Kulish states}
However, it would be consistent to define dressed states by acting with a constant dressing operator $W_\zeta$ for fixed $\zeta$ on states $\| \alpha \rrangle$,
\begin{align}
\label{eq:generalized_faddeev_kulish_states}
\| \alpha \rrangle_\zeta \equiv W^\dagger_\zeta W_\alpha \ket \alpha.
\end{align}
Physically this corresponds to defining all asymptotic states on a fixed, coherent soft boson background, defined by some momenta $\zeta$. This state does not affect the physics since soft modes decouple from Faddeev-Kulish amplitudes \cite{Mirbabayi2016} and thus this additional cloud of soft photons will just pass through the scattering process. The difference between the Faddeev-Kulish dressed state $\| \alpha \rrangle$ and the generalized states of the form $\| \alpha \rrangle_\zeta$ is that the state $\| \zeta \rrangle_\zeta = W^\dagger_\zeta W_\zeta \ket \zeta = \ket \zeta$ does not contain additional photons. This also explains why QED calculations using momentum eigenstates without any additional dressing give the correct cross-sections once we trace over soft radiation. Such a calculation can be interpreted as happening in a set of dressed states defined by
\begin{align}
\| \alpha \rrangle_{in} = W^\dagger_{in} W_\alpha \ket \alpha,
\end{align}
such that the in-state $\| in \rrangle_{in}$ does not contain photons and looks like a standard Fock-space state.

\begin{figure}[t]
\centering
\begin{subfigure}[t]{0.3\textwidth}
  \begin{tikzpicture}
  \clip (-2,-1) rectangle ++(4,4);
    \foreach \i in {0,...,5}
      \draw[line width = 1] (-2,-\i*0.2) -- ++(4,0);

    \foreach \i in {1,...,20}
      \draw[line width = 1, rotate around={45:(0,0.0)}] (\i*0.2,0) arc (0:90:\i*0.2);

    \foreach \i in {1,...,3}
      \draw[draw=yellow, snake it] (rand*0.2 ,0.1+ rand*0.2) -- (4 - rand*0.2,4- rand*0.2);
    \foreach \i in {1,...,3}
      \draw[draw=yellow, snake it] ( - rand*0.2 ,0.1+ rand*0.2) -- (-4 + rand*0.2,4- rand*0.2);
    \path[fill=gray] (-2,-0) rectangle ++(1.9,0.2);
    \path[fill=gray] (0.1,-0) rectangle ++(1.9,0.2);
  \end{tikzpicture}
    \caption{}
\end{subfigure}
\begin{subfigure}[t]{0.3\textwidth}
  \begin{tikzpicture}
  \path[fill=white] (-2,-1) rectangle ++(1,1);
    \clip (-2,1) rectangle ++(4,2);
    \path (-2,1) node[above right,opacity=1] {$\Sigma$};
    \draw [line width = 2] (-2,1) -- ++ (4,0);
    \foreach \i in {1,...,20}
      \draw[line width = 1, rotate around={45:(0,0.0)}] (\i*0.2,0) arc (0:90:\i*0.2);
    \foreach \i in {1,...,3}
      \draw[draw=yellow, snake it] (-4+ rand*0.2 ,-4+ rand*0.2) -- (4 - rand*0.2,4- rand*0.2);
    \foreach \i in {1,...,3}
        \draw[draw=yellow, snake it] (4- rand*0.2 ,-4+ rand*0.2) -- (-4 + rand*0.2,4- rand*0.2);
  \end{tikzpicture}
    \caption{}
\end{subfigure}
\begin{subfigure}[t]{0.3\textwidth}
  \begin{tikzpicture}
      \clip (-2,-1) rectangle ++(4,4);
        \foreach \i in {1,...,20}
          \draw[line width = 1, rotate =-135 ] (\i*0.2,0) arc (0:90:\i*0.2);

        \foreach \i in {1,...,20}
          \draw[line width = 1, rotate around={45:(0,0.0)}] (\i*0.2,0) arc (0:90:\i*0.2);

        \foreach \i in {1,...,3}
          \draw[draw=yellow, snake it] (-4+ rand*0.2 ,-4+ rand*0.2) -- (4 - rand*0.2,4- rand*0.2);
        \foreach \i in {1,...,3}
            \draw[draw=yellow, snake it] (4- rand*0.2 ,-4+ rand*0.2) -- (-4 + rand*0.2,4- rand*0.2);
  \end{tikzpicture}
  \caption{}
  \end{subfigure}
\caption{\textbf{(a)} A plane wave goes through a single slit and emerges as a localized wavepacket. The scattering of the incoming wavepacket results in the production of Bremsstrahlung. \textbf{(b)} We can also define some Cauchy slice $\Sigma$ and create the state by an appropriate initial condition. \textbf{(c)} Evolving this state backwards in time while forgetting about the slit results in an incoming localized particle which is accompanied by a radiation shockwave.}
\label{fig:localized_wavepacket}
\end{figure}
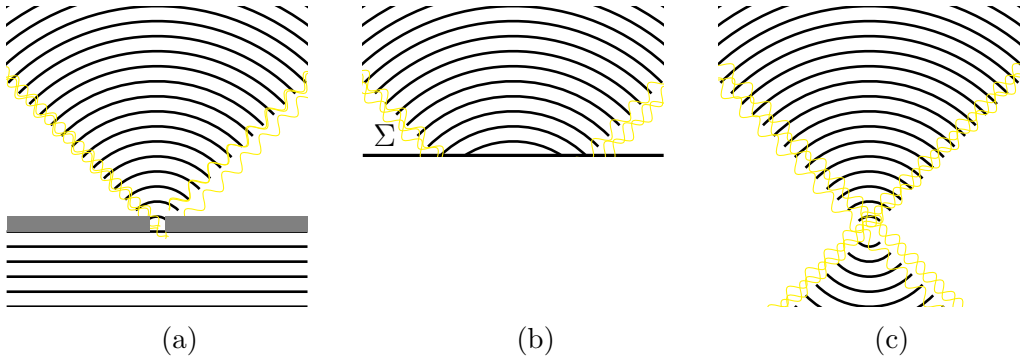

\subsubsection*{Localized particles are accompanied by radiation}
We also conclude from the previous sections that there are no charged, normalizable states which do not contain radiation. The reason is that within each selection sector there is only one non-normalizable state which does not contain radiation. Thus building a superposition to obtain a normalizable state will necessarily include dressed states which by definition contain soft bosons. A nice argument which makes this behavior plausible was given by Gervais and Zwanziger \cite{Gervais1980DerivationElectrodynamics}, see figure \ref{fig:localized_wavepacket}.

\subsection{Selection sectors}
Everything said so far has a nice interpretation in terms of the charges $\mathcal Q^\pm_{\varepsilon}$ of large gauge transformations (LGT) for QED and supertranslation for perturbative quantum gravity. For a review see \cite{Strominger2018LecturesTheory}. Large gauge transformations in QED are gauge transformations which do not die off at infinity. They are generated by an angle-dependent function $\varepsilon(\phi,\theta)$. Similarly, supertranslations in perturbative quantum gravity are diffeomorphisms which do not vanish at infinity. They are constrained by certain falloff conditions. The transformations are generated by an infinite family of charges $\mathcal Q^\pm_\varepsilon$ at future and past lightlike infinity, parametrized by a functions $\varepsilon(\phi, \theta)$ on the celestial sphere. The charges split into a hard and a soft part
\begin{align}
\mathcal Q^\pm_\varepsilon = \mathcal Q^\pm_{H,\varepsilon} + \mathcal Q^\pm_{S,\varepsilon}. 
\end{align}
The soft charge generates the transformation on zero frequency photons or gravitons and leaves undressed particles invariant, while the hard charge generates LGT or supertranslations of charged particles, i.e.~electrons in QED and all particles in perturbative quantum gravity. The action on particles can be found in \cite{Kapec2015,Campiglia2015,Gabai2016LargeQED,Choi2018AsymptoticSupertranslations}.

The charges $\mathcal Q^\pm_\varepsilon$ are conserved during time evolution (and in particular in any scattering process) and thus give rise to selection sectors of QED and gravity. These selection sectors give a different perspective on the IR catastrophe: Fock states of different momenta are differently charged under $\mathcal Q^\pm_\varepsilon$ and thus cannot scatter into each other. For dressed states, the situation is different: It was shown in \cite{Gabai2016LargeQED,Kapec2017,Choi2017} that for QED and gravity, Faddeev-Kulish dressed states $\| \alpha \rrangle$ are eigenstates of $\mathcal Q^\pm_\varepsilon$ with an eigenvalue independent of $\alpha$. 

It turns out that also our generalized version of Faddeev-Kulish states $\| \alpha \rrangle_\zeta$, equation \eqref{eq:generalized_faddeev_kulish_states}, are eigenstates of the generators $\mathcal Q^\pm_\varepsilon$ with eigenvalues which depend on $\zeta$. To see this note that \cite{Gabai2016LargeQED}
\begin{align}
[\mathcal Q^\pm_\varepsilon, W^\dagger_\zeta] = [\mathcal Q^\pm_{S,\varepsilon}, W^\dagger_\zeta] \propto \int_{S^2} d\hat{ \mathbf q} \; \frac{\zeta^2}{\zeta \cdot \hat q} \; \varepsilon(\phi,\theta) , 
\end{align}
and similarly for gravity \cite{Choi2017}. Thus the generalized Faddeev-Kulish states span a space of states which splits into selection sectors parametrized by $\zeta$. The statement that we can build physically reasonable superpositions using generalized Faddeev-Kulish states translates into the statement that superpositions can be taken within a selection sector of the LGT and supertranslation charges $\mathcal Q^\pm_\varepsilon$. 

In the context of these charges, zero energy eigenstates of $\mathcal Q^\pm_{S,\varepsilon}$ are often interpreted as an infinite set of vacua. Note that the name vacuum might be misleading as states in a single selection sector are in fact built on different vacua. Our results also raise doubt on whether physical observables exist which can take a state from one selection sector into another. If they did we could use them to create a superpositions of states from different sectors. But as we have seen above, in this case interference would not happen, which is in conflict with basic postulates of quantum mechanics.

\section{Conclusions}
\label{sec:conclusions}
Calculating cross-sections in standard QED and perturbative quantum gravity forces us to deal with IR divergences. Tracing out unobservable soft modes seems to be a physically well-motivated approach which has successfully been employed for plane-wave scattering. However, as we have shown this approach fails in more generic examples. For finite superpositions it does not reproduce interference terms which are expected; for wavepackets it predicts that no scattering is observed. We have demonstrated in this paper that dressed states \`a la Faddeev-Kulish (and certain generalizations) resolve this issue, although it is not clear if the inclusive and dressed formalism are the only possible resolutions. Importantly, we have shown that predictions of different resolutions can disagree, making them distinguishable.

Superpositions must be taken within a set of states with most of the states dressed by soft bosons. The corresponding dressing operators are only well-defined on Fock space if we use an IR-regulator which we only remove at the end of the day. In the strict $\lambda \to 0$ limit, the states are not in Fock space but rather in the much larger von Neumann space which allows for any photon content, including uncountable sets of photons \cite{Kibble1968a,Zwanziger:1974jz}. This suggests an interesting picture which seems worth investigating. The Hilbert space of QED is non-separable but has separable subspaces which are stable under action of the S-matrix and form selection sectors. These subspaces are not the usual Fock spaces but look like the state spaces defined by Faddeev and Kulish \cite{Kulish1970}, in which almost all charged states are accompanied by soft radiation. It would be an interesting task to make these statements more precise.

Our results may have implications for the black hole information loss problem. Virtually all discussions of information loss in the black hole context rely on the possibility of localizing particles -- from throwing a particle into a black hole to keeping information localized. We argued above that normalizable (and in particular localized) states are necessarily accompanied by soft radiation. It is well known that the absorption cross-section of radiation with frequency $\omega$ vanishes as $\omega \to 0$ and therefore it seems plausible that, whenever a localized particle is thrown into a black hole, the soft part of its state which is strongly correlated with the hard part remains outside the black hole. If this is true a black hole geometry is always in a mixed state which is purified by radiation outside the horizon.

\section*{Acknowledgements}
We thank Aidan Chatwin-Davies, William Donnelly, Alex May, Duff Neill, Andy Strominger, and Jordan Wilson for discussions. All of us are grateful for support from NSERC. DC was additionally supported by the Templeton Foundation and the Pacific Institute of Theoretical Physics, and DN by a UBC Four Year Doctoral Fellowship and the Simons Foundation.

\appendix

\section{Proof of positivity of $\Delta A, \Delta B$}
\label{app:formulas_exponents}
\label{app:proof_positivity}
The exponent that is responsible for the decoherence of the system is defined as 
\begin{equation}
\Delta A_{\beta\beta',\alpha\alpha'} = \frac{1}{2}A_{\beta,\alpha}+\frac{1}{2}A_{\beta',\alpha'}-\tilde{A}_{\beta\beta',\alpha\alpha'}.
\end{equation}
The factor in the first two terms, $A_{\beta,\alpha}$, is defined as in \cite{Weinberg1965}
\begin{align}
\label{eq:QEDExponent}
 A_{\beta,\alpha} & = \frac{1}{2(2\pi)^3} \int_{S^2} d\hat{\mathbf{q}} \left( \sum_{n \in \beta}\frac{e_n \eta_n p_n^\mu }{p_n \cdot \hat{{q}}}\right) g_{\mu\nu}\left(\sum_{m \in \alpha} \frac{e_m \eta_m p_m^\mu}{p_m \cdot \hat{{q}}} \right).
\end{align}
Performing the integral over $\hat{\mathbf{q}}$ yields
\begin{align}
{A}_{\beta,\alpha} = -\sum_{n,n'\in\alpha,\beta}\frac{e_{n}e_{n'}\eta_{n}\eta_{n'}}{8\pi^{2}}\beta_{nn'}\ln\left[\frac{1+\beta_{nn'}}{1-\beta_{nn'}}\right].
\end{align}
Similarly $\tilde{A}_{\beta\beta',\alpha\alpha'}$ can be written as
\begin{align}
\tilde{A}_{\beta\beta',\alpha\alpha
} = -\sum_{\substack{n\in\alpha,\beta\\n'\in\alpha'\beta'}}\frac{e_{n}e_{n'}\eta_{n}\eta_{n'}}{8\pi^{2}}\beta_{nn'}\ln\left[\frac{1+\beta_{nn'}}{1-\beta_{nn'}}\right].
\end{align}
We rearrange the terms such that $\Delta A$ can be written as 
\begin{equation}
\label{eq:Delta_A_beta_beta_alpha_alpha}
\Delta A_{\beta\beta',\alpha\alpha'} = -\frac{1}{2}\sum_{n,n'\in\alpha,\bar{\alpha}',\beta,\bar{\beta}'}\frac{e_{n}e_{n'}\eta_{n}\eta_{n'}}{8\pi^{2}}\beta^{-1}_{nn'}\ln\left[\frac{1+\beta_{nn'}}{1-\beta_{nn'}}\right],
\end{equation}
where a bar means incoming particles are taken to be outgoing and vice versa (or equivalently, $\eta_{\bar{\alpha}'} = -\eta_{\alpha'}$). From equation \eqref{eq:Delta_A_beta_beta_alpha_alpha}, it is clear that incoming particles are found within the set $\{\alpha, \beta'\}$ while the outgoing particles are part of $\{\alpha',\beta\}$. Let us rename those sets $\sigma$ and $\sigma'$ respectively. $\Delta A$ now takes the form
\begin{align}
\Delta A_{\beta\beta',\alpha\alpha'} = -\frac{1}{2}\sum_{n,n'\in\sigma,\sigma'}\frac{e_{n}e_{n'}\eta_{n}\eta_{n'}}{8\pi^{2}}\beta^{-1}_{nn'}\ln\left[\frac{1+\beta_{nn'}}{1-\beta_{nn'}}\right] = \frac{1}{2}A_{\sigma\sigma'}\geq 0, 
\end{align}
as was proven in \cite{Carney2017}. This shows that $\Delta A_{\beta\beta',\alpha\alpha'} \geq 0$. The same proof goes through for $\Delta B_{\beta\beta',\alpha\alpha'}$.

\section{The out-density matrix of wavepacket scattering}
\label{app:checks}
In this part of the appendix we flesh out the argument in section \ref{sec:continuous}, namely that after tracing out soft radiation, the only contribution to the out-density matrix is coming from the identity term in the S-matrix. We will focus on the case of QED.
\subsection{Contributions to the out-density matrix}
First, let us decompose the IR regulated S-matrix into its trivial part and the $\mathcal M$-matrix element. For simplicity we ignore partially disconnected terms, where only a subset of particles interact. Then,
\begin{align}
S^\Lambda_{\alpha \beta} = \delta(\alpha- \beta) - 2 \pi i \mathcal M^\Lambda_{\alpha \beta} \delta^{(4)}(p^\mu_\alpha - p^\mu_\beta),
\end{align}
where the first term is the trivial LSZ constribution to forward scattering. This trivial part does not involve any divergent loops and therefore exhibits no $\Lambda$-dependence. However, the factorization of the S-matrix into a cutoff dependent term times some power of $\lambda/\Lambda$ remains valid since all exponents of the form $A_{\alpha,\beta}$ vanish identically for forward scattering. This decomposition of the S-matrix gives rise to three different terms for the outgoing density matrix, containing different powers of $\mathcal M$.
\subsubsection*{``No scattering''-term}
The case where both S-matrices contribute the delta function term results -- unsurprisingly -- in the well-defined outgoing density matrix
\begin{align}
\rho^{(I)}_{\beta\beta'} = \int d \alpha d \alpha'  f(\alpha) f(\alpha')^* \delta(\alpha- \beta) \delta(\alpha'- \beta') \delta_{\alpha \alpha'} = f(\beta) f^*(\beta').
\end{align}
\subsubsection*{Contribution from forward scattering}
We would now expect to find an additional contribution to the density matrix reflecting the non-trivial scattering processes, coming from the cross-terms
\begin{align}
  \label{eq:integrand_rho_II}
  - 2 \pi i \left( \delta(\alpha - \beta) \mathcal M^\Lambda_{\alpha' \beta} \delta^{(4)}(p^\mu_{\alpha'} - p^\mu_\beta) - \delta(\alpha' - \beta) \mathcal M^{\dagger \Lambda}_{\alpha \beta} \delta^{(4)}(p^\mu_\alpha - p^\mu_\beta)\right).
\end{align}
For simplicity, let us focus solely on the case in which $S^{*}$ contributes the delta function and $S$ contributes the connected part
\begin{align}
\rho^{(II)}_{\beta\beta'} = - 2 \pi i f^*(\beta')  \int d\alpha f(\alpha)  \mathcal M^\Lambda_{\beta \alpha} \delta^{(4)}(p^\mu_\alpha - p^\mu_\beta) \lambda^{\Delta A_{\alpha,\beta}} \mathcal F(E,E_T, \Lambda)_{\beta,\alpha} + \dots,
\end{align}
where the ellipsis denotes the contribution coming from the omitted term of \eqref{eq:integrand_rho_II}. The exponent of $\lambda$ only vanishes if the currents in $\alpha$ and $\beta$ agree. We will show in appendix \ref{sec:exchangeoflimits} that we can take the limit $\lambda \to 0$ before doing the integrals. Taking this limit, $\lambda^{\Delta A_{\alpha,\beta}}$ gets replaced by
\begin{align}
\delta_{\alpha\beta} = \begin{cases} 1, \text{ if charged particles in $\alpha$ and $\beta$ have the same velocities } \\ 0 ,\text{  otherwise},\end{cases}
\end{align}
which is zero almost everywhere. If the integrand was regular, we could conclude that the integrand is a zero measure subset and integrates to zero and thus
\begin{align}
	\label{eq:density_matrix_II}
	\rho^{(II)}_{\beta\beta'} = 0.
\end{align}
However, the integrand is not well-behaved. Singular behavior can come from the delta function or the matrix element, so let's consider the two possibilities. 

The singular nature of the Dirac delta does not affect our conclusion: for $n$ incoming particles, the measure $d\alpha$ runs over $3n$ momentum variables while the delta function constrains $4$ of them, leaving us with $3n-4$ independent ones. If we managed to find a configuration for which $\Delta A_{\beta\alpha} = 0$, any infinitesimal variation of the momenta in $\alpha$ along a direction that conserves energy and momentum would modify the eigenvalue of the current operator $\hat j_{v}(\alpha)-\hat j_{v}(\beta)$ and make $\Delta A_{\beta\alpha}$ non-zero. Therefore, the integrand would still be a zero-measure subset for the remaining integrals.

What could still happen is that $\mathcal M^\Lambda_{\beta\alpha}$ is so singular that it gives a contribution. For this to happen it would need to have contributions in the form of Dirac delta functions. However, also this does not happen, for example for Compton scattering which scatters into a continuum of states. Additional IR divergences also do not appear as guaranteed by the Kinoshita-Lee-Nauenberg theorem. We will not give a general proof since for our purposes it is problematic enough to know that no scattering is observed for \emph{some} physical process.

\subsubsection*{The scattering term}
It is evident that a similar argument goes through for the $\mathcal M^2$ term. One finds
\begin{align}
\rho^{(III)}_{\beta\beta'} = -4 \pi^2 &\int d\alpha d\alpha' f(\alpha) f^*(\alpha')  {\mathcal M}^\Lambda_{\beta \alpha}{\mathcal M}^{\Lambda*}_{\alpha' \beta'}\lambda^{\Delta A_{\alpha \alpha',\beta \beta'}} \\
& \times  \mathcal F(E,E_T, \Lambda)_{\beta\beta',\alpha\alpha'} \delta^{(4)}(p^\mu_\alpha - p^\mu_\beta) \delta^{(4)}(p^\mu_{\alpha'} - p^\mu_{\beta'}).
\end{align}
The analysis boils down the the question whether the term
\begin{align}
\int d\alpha d\alpha' \lambda^{\Delta A_{\alpha \alpha',\beta \beta'}} \delta^{(4)}(p^\mu_\alpha - p^\mu_\beta) \delta^{(4)}(p^\mu_{\alpha'} - p^\mu_{\beta'}).
\end{align}
vanishes. As soon as there is at least one particle with charge, we need to obey the condition that the charged particles in $\alpha$ and $\beta'$ agree with those in $\beta$ and $\alpha'$ for the exponent of $\lambda$ to vanish. Infinitesimal variations of $\alpha$ and $\alpha'$ that preserve the eigenvalue of the current operator $\hat j_{v}(\alpha) - \hat j_{v}(\alpha')$ form a zero-measure subset of the $6n-8$ directions that preserve momentum and energy, forcing us to conclude that the integration runs over a zero measure subset and the only contribution to the reduced density matrix comes from the trivial part of the scattering process. This means that
\begin{align}
\rho^{out, red.}_{\beta\beta'} = f(\beta) f^*(\beta') = \rho^{in}_{\beta\beta'},
\end{align}
or in other words it predicts that a measurement will not detect scattering for wavepackets. This is clearly in contradiction with reality and suggests that the standard formulation of QED and perturbative quantum gravity which relies on the existence of wavepackets is invalid.

\subsection{Taking the cutoff $\lambda \to 0$ vs. integration}
\label{sec:exchangeoflimits}
One might be concerned that the limit $\lambda \to 0$ and the integrals do not commute. In this part of the appendix, we will check the claim made in the preceding subsection, i.e. we will show that one can explicitly check that the integration and taking the IR regulator $\lambda$ to zero commute. 
We assume in the following that we talk about QED with electrons and muons in the non-relativistic limit, which again is good enough as it is sufficient to show that we can find a limit in which no sign of scattering exists in the outgoing hard state. The wave packets are chosen to factorize for every particle and to be Gaussians in velocity centered around $v=0$,
\begin{align}
f(v) = \left(\frac{2}{\pi  \kappa }\right)^{3/4} \exp \left(-\frac{v^2}{\kappa }\right).
\end{align}
In order to stay in the non-relativistic limit, $\kappa$ must be sufficiently small. They are normalized such that
\begin{align}
\int d^3v |f(v)|^2 = 1.
\end{align}
In the exponent of $\lambda$ we set $\alpha' = \beta'$ for simplicity, i.e.~we consider the case of forward scattering. In the non-relativistic limit, we can expand the exponent of $\lambda$ into
\begin{align}
\Delta A_{\alpha\beta} = \frac{e^2}{24 \pi^2} \sum_{n,m \in \alpha, \beta} (v_\alpha - v_\beta)^2.
\end{align}
Thus, $\lambda^{\Delta A}$ has the form
\begin{align}
\lambda^{\Delta A} \propto \exp\left(- \frac 1 2 \gamma \sum_{n,m \in \alpha, \beta} (v_\alpha - v_\beta)^2 \right),
\end{align}
where taking the cutoff $\lambda$ to zero corresponds to $\gamma \propto - \log(\lambda) \to \infty$. The state $\alpha$ consists of a muon with well defined momentum and one electron with momentum $m v$, where $v$ is centered around $0$. The state $\beta$ consists of the same muon (we assume it was not really deflected) and one electron with momentum $m v'$. To obtain the contribution to forward scattering, we have to perform the integral
\begin{align}
\propto \int d^3v \left(\frac{2}{\pi \kappa }\right)^{3/4} \exp\left(-\frac{v^2}{\kappa }\right) \exp\left(- \gamma (v - v')^2\right) \cdot \text{(other terms)}.
\end{align}
Here, we assumed that the other terms which include the matrix element in the regime of interest is finite and approximately independent of $v$. The integral yields
\begin{align}
\left( \frac{2 \pi \kappa}{(1 + \gamma \kappa)^2}\right)^{3/4}\exp\left({-\frac{\gamma v'^2}{1 + \gamma \kappa}}\right).
\end{align}
Taking the limit $\gamma \to \infty$, it is clear that this expression vanishes. If we want to consider an outgoing wave packet we have to integrate this over $f(v' - v_{out})$. The result is proportional to
\begin{align}
\left( \frac{2 \pi \kappa}{(1 + 2 \gamma \kappa)^2}\right)^{3/4}\exp\left({-\frac{\gamma v_{out}^2}{1 + 2 \gamma \kappa}}\right)
\end{align}
and still vanishes if we remove the cutoff, $\gamma \to \infty$.

\bibliographystyle{utphys-dan}
\bibliography{Mendeley_IR_Divergences}

\end{document}